\documentclass[10pt]{article} 
\usepackage[margin=1in,headsep=.2in]{geometry}
\usepackage{amsmath, amsfonts, amssymb, amsthm}
\usepackage{graphicx, float, caption, subcaption}
\usepackage{multirow}
\usepackage{algorithm}
\usepackage[title]{appendix}
\usepackage{authblk}
\usepackage{booktabs}

\usepackage[numbers]{natbib}
\usepackage[group-separator={,}, group-minimum-digits={3}]{siunitx}

\theoremstyle{definition}
\newtheorem*{example*}{Example}

\title{Finding and assessing treatment effect sweet spots in clinical trial data}
\author[1]{Erin Craig}
\author[2]{Donald A Redelmeier}
\author[3]{Robert J Tibshirani}
\affil[1]{Department of Biomedical Data Sciences, Stanford University, Stanford, CA, USA.}
\affil[2]{Department of Medicine, University of Toronto, Toronto, Ontario, Canada; Evaluative Clinical Sciences Department, Sunnybrook Research Institute, Toronto, Ontario, Canada; Population and Global Health Department, Institute for Clinical Evaluative Sciences, Toronto, Ontario, Canada; Division of General Internal Medicine, Sunnybrook Health Sciences Centre, Toronto, Ontario, Canada; Center for Leading Injury Prevention Practice Education \& Research, Toronto, Ontario, Canada.}
\affil[3]{ Department of Biomedical Data Sciences, Stanford University, Stanford, CA, USA; Department of Statistics, Stanford University, Stanford, CA, USA.}

\begin{document}
\bibliographystyle{unsrt} 

\maketitle

\begin{abstract}
Identifying heterogeneous treatment effects (HTEs) in randomized controlled trials is an important step toward understanding and acting on trial results. However, HTEs are often small and difficult to identify, and HTE modeling methods which are very general can suffer from low power. We present a method that exploits any existing relationship between illness severity and treatment effect, and identifies the ``sweet spot", the contiguous range of illness severity where the estimated treatment benefit is maximized. We further compute a bias-corrected estimate of the conditional average treatment effect (CATE) in the sweet spot, and a $p$-value. Because we identify a single sweet spot and $p$-value, we believe our method to be straightforward to interpret and actionable: results from our method can inform future clinical trials and help clinicians make personalized treatment recommendations. 
\end{abstract}
%

\section{Introduction}

Randomized trials often need large sample sizes to achieve adequate statistical power, and recruiting patients across the spectrum of illness severity can make it easier to find a sufficiently large cohort. However, this recruitment strategy can be a double-edged sword, as patients with mild or severe illness may receive little benefit from treatment~\cite{redelmeier2020approach}. For patients with mild illness, treatment may be superfluous; for patients with severe illness, treatment may be futile. Including these patients in a study may bias a study toward the null, as the study results rely on a subset of patients with illness severity in the middle range. 

We present a simple approach that identifies the single contiguous range of illness severity where the estimated treatment benefit is maximized. We consider this range to be the ``sweet spot" for treatment. We further present methods to compute a bias-corrected estimate of the conditional average treatment effect (CATE), and to control type I error.  Because we identify a single sweet spot and compute a $p$-value, we believe our method to be straightforward to interpret and actionable: results from our method can inform future clinical trials and help clinicians make personalized treatment recommendations.  

As a running example to illustrate our method, we use the AQUAMAT trial~\cite{dondorp2010artesunate}, which compares artesunate to quinine in the treatment of severe falciparum malaria in African children. This randomized trial studied $\num{5488}$ children with severe malaria, with primary outcome in-hospital mortality. This study was conducted between Oct 3, 2005, and July 14, 2010. Half of the patients were randomized to receive artesunate, and half to receive quinine. The trial found that artesunate substantially reduces mortality in African children with severe malaria.  The patients in this study were diverse across $45$ measured covariates including age, sex, complications on admission, vitals, and labs (for a full description of covariates, we refer to \cite{dondorp2010artesunate}). This diversity makes it more likely that some patients would do well or poorly regardless of care, though it is not obvious how to identify them: a Mantel-Haenszel subgroup analysis showed no evidence of any differences in outcomes between subgroups. However, the reanalysis of this trial in \cite{watson2020graphing} suggests that there may be treatment effect heterogeneity along the axis of illness severity.

\begin{figure}[H]
\centering
  \includegraphics[width=.6\linewidth]{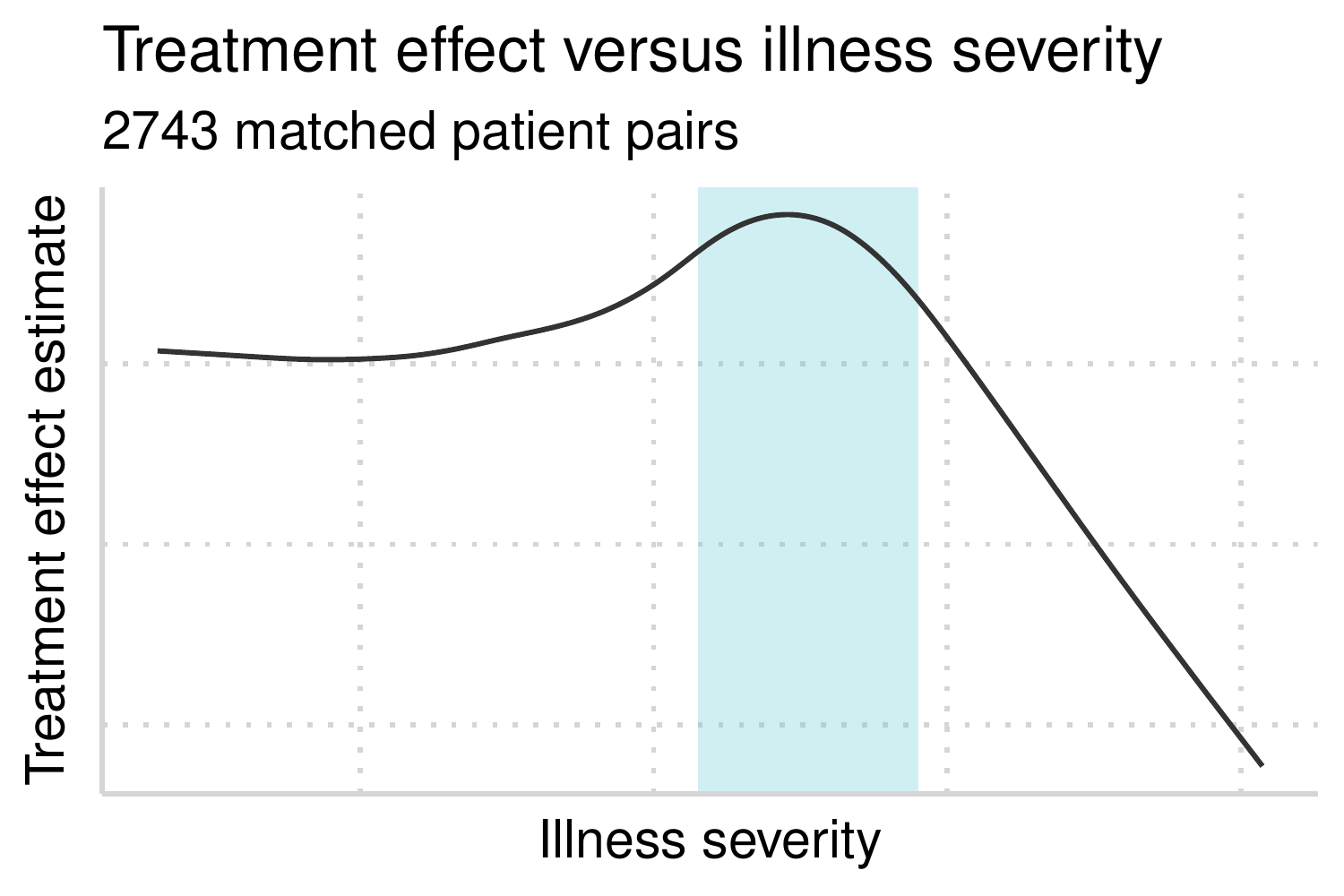}
  \caption{Smoothed treatment effect as a function of illness severity on the AQUAMAT randomized trial. By visual inspection, it seems that patients with illness severity in the shaded range seem to benefit more from treatment.}
  \label{fig:intro}
\end{figure}

In Figure \ref{fig:intro}, we visualize the smoothed treatment effect estimate as a function of illness severity for the patients in this trial. From this image, it is tempting to determine that there is a range of illness severity where patients receive more benefit from treatment; acting on this determination can be dangerous, however, as the apparent sweet spot could be due to chance alone. Our statistical framework protects against this by finding sweet spots and judging their significance.

\section{Related work}

There has been recent interest in developing methods to estimate heterogeneous treatment effects in randomized trials \cite{redelmeier2020approach, zhao2013effectively, athey2016recursive, athey2019generalized, kunzel2019metalearners, watson2020machine, watson2020graphing}. 

Athey and Imbens~\cite{athey2016recursive} model treatment effect as a function of patient covariates: the causal tree estimator is a tree that partitions the covariate space into subsets where patients have lower-than or higher-than-average treatment effects, and allows valid inference for causal effects in randomized trials. Causal trees can be combined in an ensemble to form a causal forest~\cite{athey2019generalized} which is more flexible, though harder to interpret. Instead of modeling treatment effect, Watson and Holmes~\cite{watson2020machine} identify the presence of treatment effect heterogeneity with control of type I error. This method uses a statistical test to compare treated and untreated outcomes among a subgroup of patients predicted to benefit from treatment. This is repeated many times on different subsets of the data, and the tests are summarized into a single $p$-value that accounts for multiple hypothesis testing. As in \cite{athey2016recursive}, this method may uncover treatment effect heterogeneity even when the relationship between covariates and the outcome is complex. 

However, heterogeneous treatment effects are often small and difficult to identify, and methods which are very general can suffer from low power. Rather than search the full covariate space, the methods in Redelmeier and Tibshirani~\cite{redelmeier2020approach}, Watson and Holmes~\cite{watson2020graphing} and Kent et al~\cite{kent2010assessing} look directly at the relationship between a precomputed measure of illness severity and treatment effect. The method in \cite{redelmeier2020approach} orders patients by increasing illness severity, computes the cumulative treatment effect, and compares the goodness of fit of a line and a Gompertz CDF -- there is no heterogeneity when the cumulative treatment effect is linear. The method in \cite{watson2020graphing} models individual treatment effect as a function of illness severity using a localised reweighting kernel. Finally, the method in~\cite{kent2010assessing} stratifies patients by risk, and then estimates the treatment effect separately on each stratum. None of these methods quantify type I error or statistical power. 

\section{Our proposed method}
Our method exploits any existing relationship between between illness severity and treatment effect, and searches for the contiguous range of illness severity where the estimated treatment benefit is maximized. We further compute a bias-corrected estimate of the conditional average treatment effect (CATE) in the sweet spot, and compute a $p$-value. 

\begin{algorithm}
  \caption{Finding and assessing a sweet spot on clinical trial data\\for a randomized trial design with a control$:$treated ratio of  $k$:$1$}
  \begin{enumerate}
    \item Compute a predilection score for each patient that indicates illness severity.
    \item Create sets of patients with similar scores, consisting of $k$ controls and one treated patient. On each matched set, compute the treatment effect, the difference between the treated outcome and average control outcome.
    \item Perform an exhaustive search to identify the sweet spot --- the range of illness severity scores where the treatment effect is maximal.
    \item Test the null hypothesis that there is no sweet spot related to illness severity with a permutation test.
    \item Debias the estimate of the treatment effect inside and outside the sweet spot using the parametric bootstrap.
  \end{enumerate}
  \label{alg:overview}
\end{algorithm}

\subsection{Computing predilection scores} 
Illness severity is characterized by prognosis at baseline: we refer to this as a \textit{predilection score}, as it represents the patient's baseline predilection to the outcome. Predilection scores are computed from a model trained to predict the outcome from the baseline covariates, and they may take on any real values. To model continuous or binary outcomes, we recommend linear or logistic regression. 

There are two important considerations when fitting the predilection score model. First, the model must be trained only on the control patients, as the intervention may have altered the natural history of the treated patients. Second, prevalidation must be used to avoid overfitting to the controls~\cite{tibshirani2002pre, abadie2018endogenous}; prevalidation ensures that every patient's prediction comes from a model trained solely on other patients. To do $k$-fold prevalidation, partition the controls evenly into $k$ sets, train a predilection score model on $k-1$ sets and use this model to compute scores on the remaining set. This is repeated so that every set is held out, and as a result, no patient is used to train the model that ultimately computes their predilection score. We illustrate this on a small example in Table~\ref{table:preval}. We thank Lu Tian, Trevor Hastie and Rocky Aikens for bringing this to our attention, and we further motivate the need for prevalidation in Section~\ref{section:preval}. 

We note that it may not be necessary to train a new predilection score model when there already exists a model trained on an external dataset of patients who received the ``control'' treatment. 

\begin{table}[H]
\begin{center}
\begin{tabular}{ l | l }
train & predict\\
\hline
 controls 3--10 &  controls 1, 2  \\  
 controls 1, 2, 5--10 & controls 3, 4  \\ 
 controls 1--4, 7--10 &  controls 5, 6  \\ 
 controls 1--6, 9, 10 &  controls 7, 8  \\ 
 controls 1--8 &  controls 9, 10  \\  
\end{tabular}
\caption{An example of five-fold prevalidation on $10$ controls. More generally, when doing $k$-fold prevalidation, we use $k$ models, each trained on $\frac{(k-1)n}{k}$ controls and used to compute the score on the remaining $\frac{n}{k}$ controls.}
\label{table:preval}
\end{center}
\end{table}

\begin{figure}[H]
  \begin{minipage}[b]{0.5\linewidth}
      \centering
  \begin{tabular}{ l  r  }
& odds ratio \\
\hline
intercept& $0.080$\\
coma& $4.765$\\
shock & $1.563$\\
convulsions & $1.371$\\
base deficit & $0.986$\\
BUN & $1.016$\\
hematocrit & $0.995$\\
respiratory rate & $1.013$\\
bicarbonate & $0.874$\\
hypoglycaemia & $2.410$\\
\hline
\end{tabular}
 \par\vspace{0pt}
   \end{minipage}%
  \begin{minipage}[b]{0.48\linewidth}
 \centering
   \includegraphics[width=\linewidth]{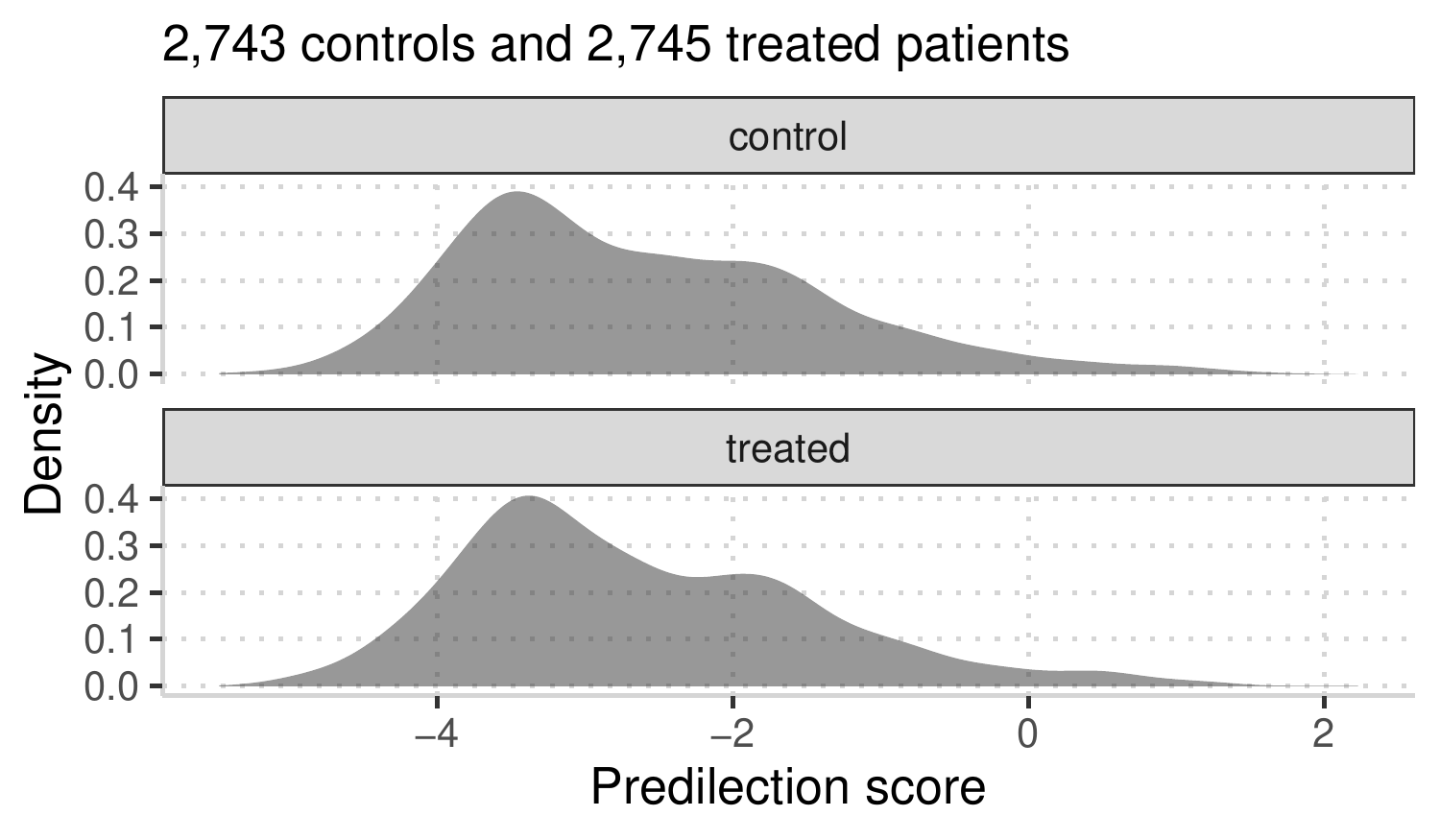}
   \par\vspace{0pt}
\end{minipage}
\caption{Odds ratios for the predilection score model, and the distributions of predilection scores for treated and control patients. An odds ratio above $1$ indicates increased risk of death.}
\label{fig:logreg}
\end{figure}

\begin{example*}
To compute predilection scores on the AQUAMAT data, we choose logistic regression: our outcome is in-hospital mortality. We begin with $10$-fold prevalidation to compute scores for the $\num{2743}$ control patients in the trial, and then we fit a model on all $\num{2743}$ controls and compute the predilection scores of the treated patients. The predilection scores have moderate goodness-of-fit (AUROC = $0.82$, AUPRC = $0.78$), and we report the odds ratios in Figure~\ref{fig:logreg}. We also illustrate the distribution of scores for the treated and control patients.
\end{example*}

\subsection{Estimating treatment effects}
\label{section:treatmenteffects}

We now estimate treatment effect as a function of predilection score. For a trial design with a control$:$treated ratio of $k$:$1$, we use optimal matching~\cite{hansen2019package} to form groups of $k$ controls and one treated patient with similar predilection scores. Each group's predilection score is their average predilection score, and for binary or continuous outcomes, their conditional average treatment effect (CATE) is the mean difference between the treated and control outcomes within the group.

\begin{figure}[H]
\centering
\begin{subfigure}{.5\linewidth}
  \centering
  \includegraphics[width=.9\linewidth]{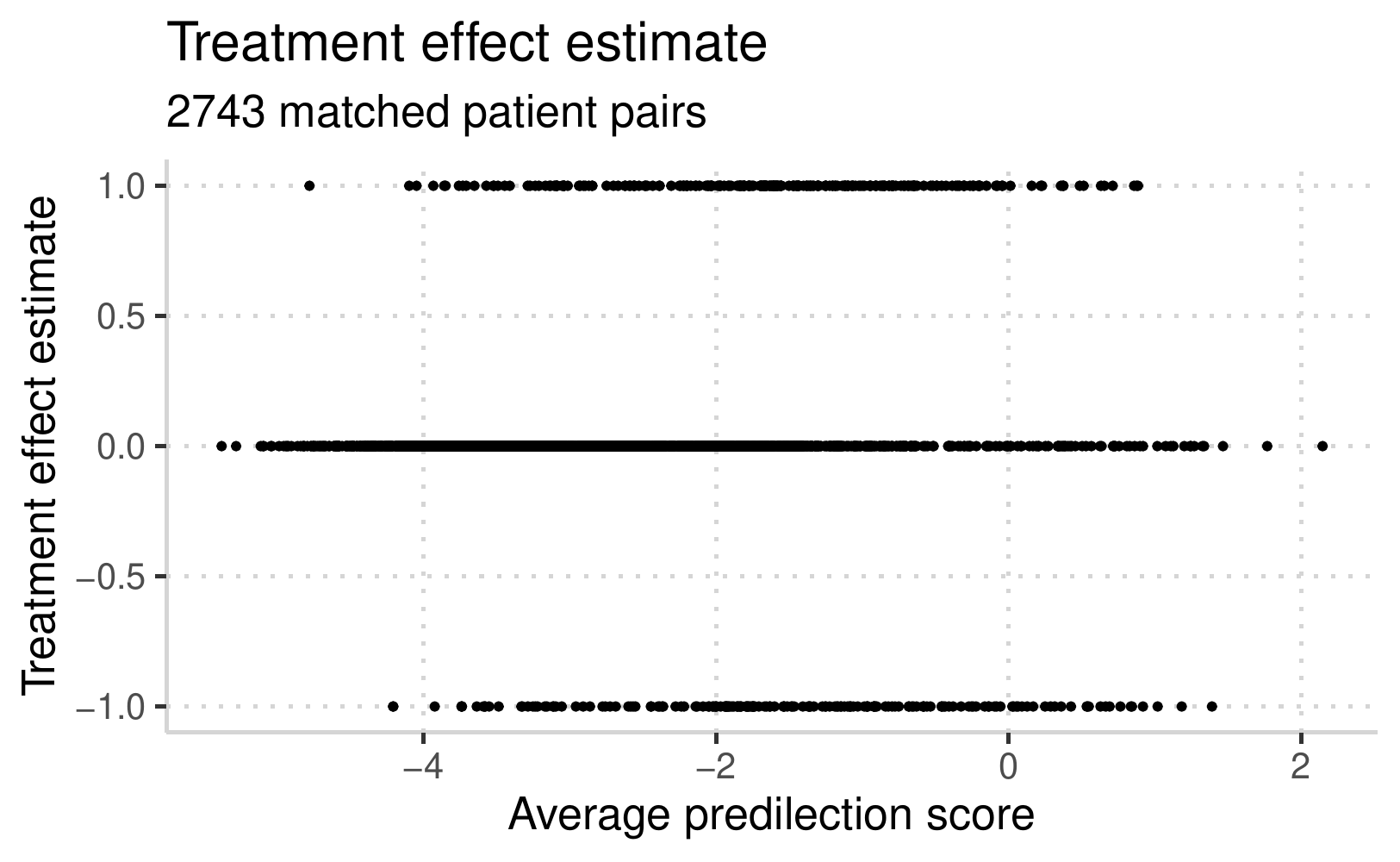}
\end{subfigure}%
\begin{subfigure}{.5\linewidth}
  \centering
  \includegraphics[width=.9\linewidth]{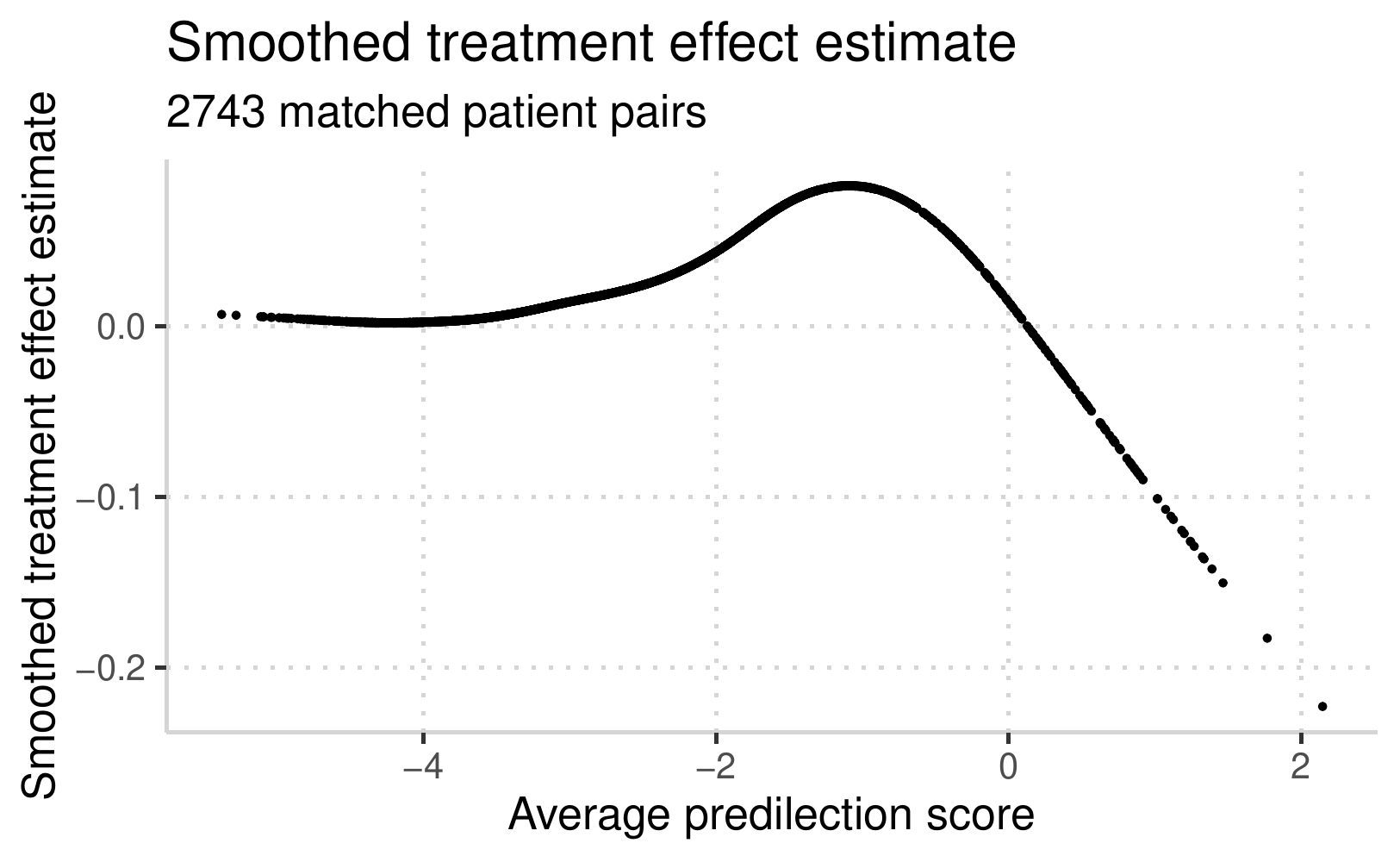}
\end{subfigure}
\caption{On the left, treatment effect as a function of illness severity on the AQUAMAT randomized trial. On the right, a smoothing spline fit to the treatment effects suggests a region of predilection scores where patients respond more strongly to treatment.}
\label{fig:matched_triplets}
\end{figure}

\begin{example*}
On the AQUAMAT data, we have a control$:$treated ratio of $1$:$1$, and we form $\num{2743}$ sets containing one control and one treated patient with similar predilection scores. On each matched set, we compute the treatment effect as the difference in in-hospital mortality between the treated and control patient. For example matched sets and their estimated treatment effects, see Table~\ref{table:examplescores}; for the treatment effect as a function of predilection score, see Figure~\ref{fig:matched_triplets}. 
\end{example*}

\begin{table}[H]
\centering
\begin{tabular}{ r r |  r r | r  r }
\multicolumn{2}{c }{predilection score} & \multicolumn{2}{c }{in-hospital mortality} &  & \\
\hline
\multicolumn{1}{c}{control} & \multicolumn{1}{c}{treated} & \multicolumn{1}{c}{control}  & \multicolumn{1}{c}{treated} & \multicolumn{1}{c}{mean score} & \multicolumn{1}{c}{CATE} \\
\toprule
$-4.21$ & $-4.21$ & $0$ & $1$ & $-4.21$ & $-1$\\
\rule{0pt}{2.6ex}
$-5.34$ & $-5.38$ & $0$ & $0$ & $-5.36$ & $0$\\
\rule{0pt}{2.6ex}
$-1.98$ & $-1.97$ & $1$ & $1$ & $-1.97$ & $0$\\
\rule{0pt}{2.6ex}
$-4.78$ & $-4.78$ & $1$ & $0$ & $-4.78$ & $1$\\
\end{tabular}
\caption{Predilection scores, outcomes and estimated treatment effects for four example matched sets in the AQUAMAT data. A lower predilection score indicates less severe illness.}
\label{table:examplescores}
\end{table}

\subsection{Finding the sweet spot}
 \label{section:sweetspot} 
We have defined $\mathbf{t} = \{t_k\}_{k=1}^n$ and $\mathbf{s} = \{s_k\}_{k=1}^n$, the sequences of estimated treatment effects and predilection scores of our matched sets, both ordered by increasing predilection score. That is, $s_1 \leq s_i \leq s_n$, and $t_i$ is the treatment effect for the set with predilection score $s_i$.

We would like to identify a contiguous subsequence of $\mathbf{t}$ that (1) is long (to cover as many patients as possible), and (2) has a large average treatment effect. To measure the extent to which any subsequence meets our requirements, we use the length of the sequence (which captures criterion 1) times the difference between the sequence average and the global average (which captures criterion 2). Explicitly, for the subsequence of $\mathbf{t}$ consisting of $\{t_i, t_{i+1}, \dots, t_j\}$, compute:
\[
	Z(i, j) = (j-i+1)  \left( \text{mean}\left(\{t_k\}_{k=i}^j\right) - \text{mean}\left(\{t_k\}_{k=1}^n\right) \right).
\]
 
The values of $i$ and $j$ that maximize $Z$ indicate the location of the sweet spot, and they are found by an exhaustive search over $i \in [1, n-1],\, j \in [i+1, n]$. We write these values as $(\hat{i}, \hat{j}) = \arg \max_{i, j} Z(i, j)$; the sweet spot includes patients with predilection scores between $s_{\hat{i}}$ and $s_{\hat{j}}$. 

\begin{figure}[H]
\centering
  \includegraphics[width=.8\linewidth]{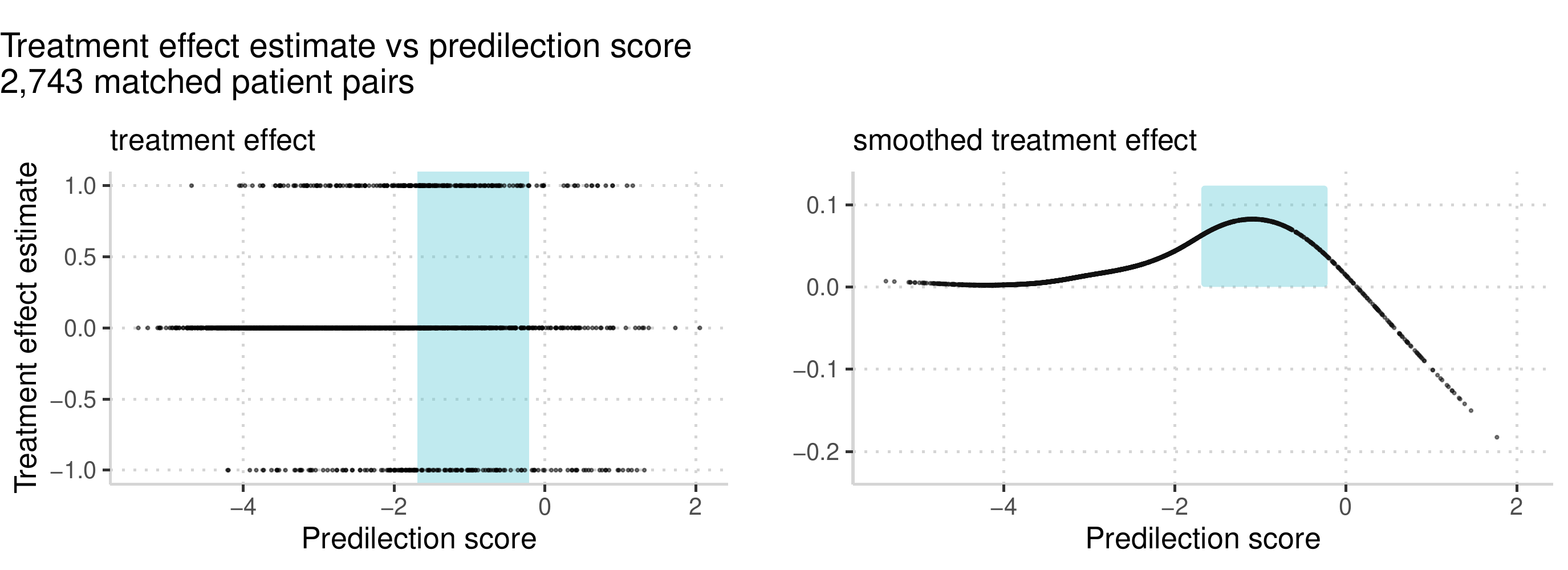}
 \caption{The sweet spot identified on the AQUAMAT data. On the left, we highlight the range of predilection scores in the sweet spot. On the right, we show the mean treatment effect estimate inside the sweet spot. For illustration, we include a smoothing spline fit to the treatment effect estimate.}
  \label{fig:sweet_spot_1}
\end{figure}

\begin{example*} On the AQUAMAT data, the maximum value of $Z$ is $\widehat{Z} = 47.47$, with sweet spot $(\hat{i}, \hat{j}) = (2153, 2631)$  corresponding to patients with predilection scores between $-1.70$ and $-0.20$. The mean treatment effect in this range is $0.12$; outside this range, it is $0.00$. This is illustrated in Figure~\ref{fig:sweet_spot_1}.
\end{example*}

\subsection{Calibrating}
\label{section:pvalue}

%
%

We wish to test the null hypothesis that there is no sweet spot related to illness severity. We ask: if there were \textit{no sweet spot}, how often would we observe a value at least as large as $\widehat{Z} = Z(\hat{i}, \hat{j})$?

Suppose there is no sweet spot, and the true treatment benefit is the same across the entire range of illness severity. In this case, the ordering of the treatment effect sequence $\mathbf{t}$ does not matter: with the same probability, we could have observed any permutation of $\mathbf{t}$, and the maximum value of $Z$ on the permuted sequence would be similar to $\widehat{Z}$. However, if there is a sweet spot, the ordering of $\mathbf{t}$ is meaningful, and $\widehat{Z}$ would be larger than most of the maximum values of $Z$ on the permuted sequences.

We test our null hypothesis with a permutation test: we repeatedly shuffle the values of $\mathbf{t}$ and find the maximum value of $Z$ on the permuted sequence. The $p$-value is the relative frequency that the maximum value on the permuted sequence is at least as large as $\widehat{Z}$. 

\begin{figure}[H]
  \centering
  \includegraphics[width=.5\linewidth]{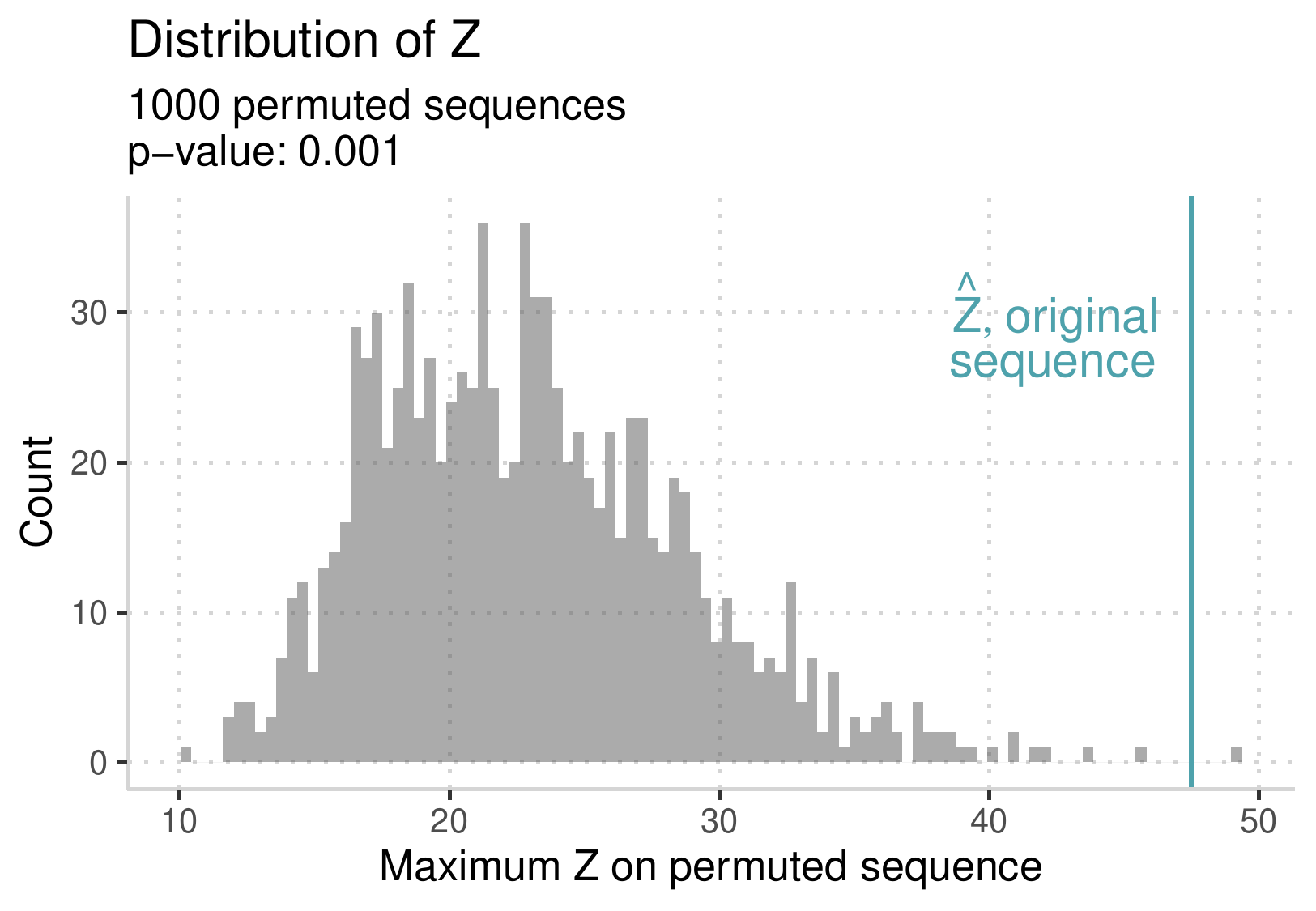}
\caption{Sweet spot permutation test on the AQUAMAT data. 
}
\label{fig:pvalue}
\end{figure}

\begin{example*} We do a permutation test on the AQUAMAT data. In Section~\ref{section:treatmenteffects}, we computed the ordered sequence of treatment effects, and in Section~\ref{section:sweetspot} we chose the sweet spot corresponding to $\widehat{Z} =  47.47$. For \num{1000} iterations, we permuted the sequence of treatment effects and computed the maximum value of $Z$. On one of those permutations, we observed a value of $Z$ that was larger than $47.47$, which corresponds to $p$-value $0.001$. We visualize our permutation test in Figure~\ref{fig:pvalue}. 
\end{example*}

\subsection{Debiasing}

Finally, we wish to estimate the treatment effect in the sweet spot. The naive choice is the mean treatment effect in the sweet spot: $\hat{\tau} = \text{mean}(\{t_{\hat{i}}, \dots, t_{\hat{j}}\})$. However, this may be optimistic, as searching for the sweet spot may bias the treatment effect. We debias our estimate with the parametric bootstrap~\cite{tibshirani1993introduction}, using our sweet spot as a model to simulate data. 

Having computed treatment effects $\mathbf{t} = \{t_k\}_{k=1}^n$ and a sweet spot location $[\hat{i}, \hat{j}]$, we generate a new sequence of treatment effects $\mathbf{t}^*$. The values inside the sweet spot $\{t^*_k\}_{k=\hat{i}}^{\hat{j}}$ are sampled with replacement from  $\{t_k\}_{k=\hat{i}}^{\hat{j}}$, the values inside the sweet spot on the original sequence. Likewise, the values outside the sweet spot, $\{t^*_k\}_{k=1}^{\hat{i}-1} \cup \{t^*_k\}_{k=\hat{j}+1}^{n}$, are sampled with replacement from $\{t_k\}_{k=1}^{\hat{i}-1} \cup \{t_k\}_{k=\hat{j}+1}^{n}$.  We repeatedly simulate data using this method, find its sweet spot, and estimate the CATE in the sweet spot. Our bootstrapped bias estimate is the difference between the mean bootstrapped CATE estimate, $\hat{\tau}_\text{boot}$, and $\hat{\tau}$. To bias-correct $\hat{\tau}$, subtract the bias: 
\begin{align*}
\hat{\tau}_{\text{corrected}} &= \hat{\tau}- \widehat{\text{bias}}\\
&= \hat{\tau}- \left(\hat{\tau}_\text{boot} - \hat{\tau}\right)\\
&= 2\, \hat{\tau}- \hat{\tau}_\text{boot}.
\end{align*}

In every run of the bias-correction bootstrap, we also obtain an estimate of the location of the sweet spot. These estimates form an empirical distribution around the values of $\hat{i}$ and $\hat{j}$ in the original estimation of the sweet spot.

\begin{figure}[H]
\centering
\begin{subfigure}{.5\linewidth}
    \includegraphics[width=.9\linewidth]{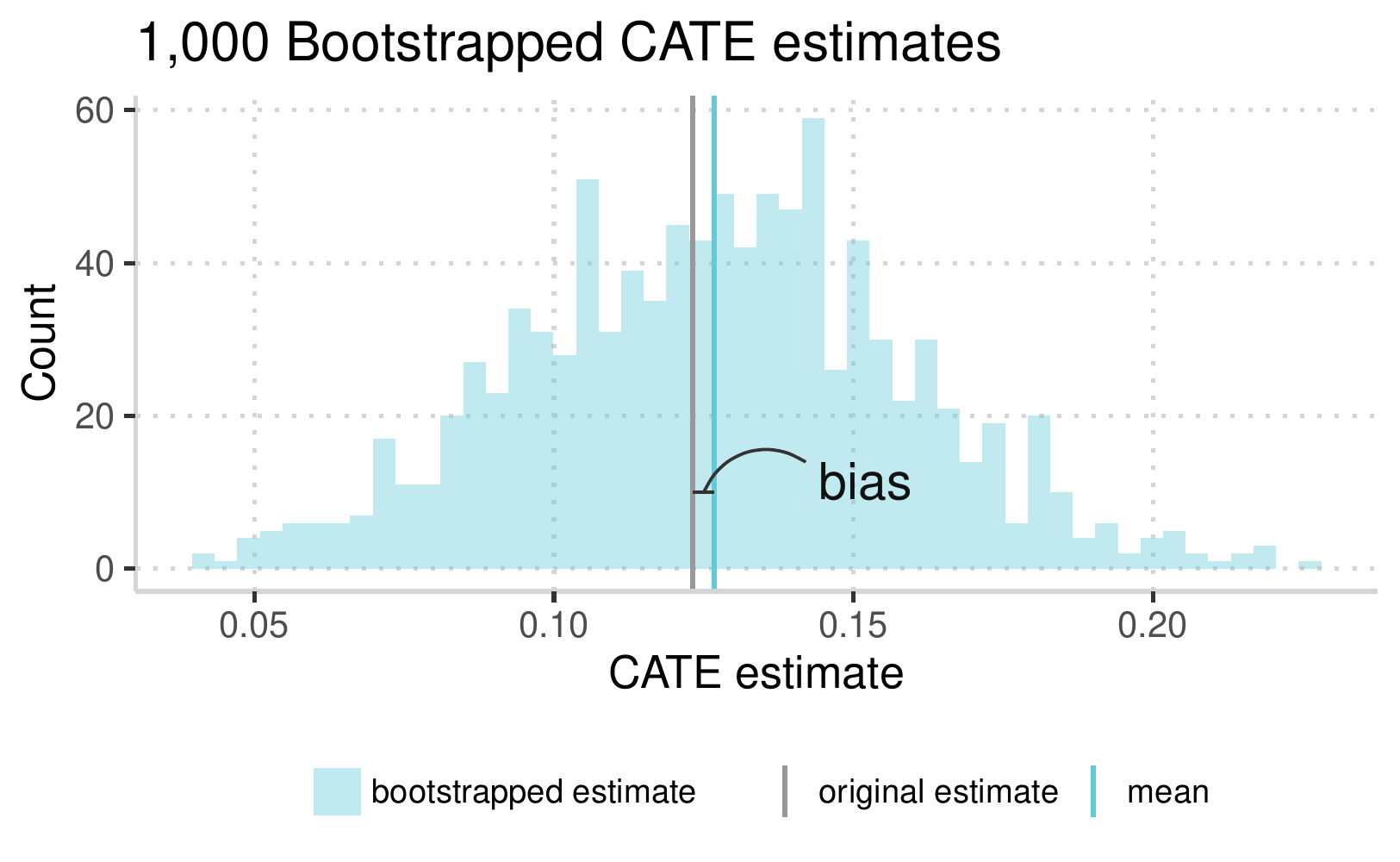}
\end{subfigure}%
\begin{subfigure}{.5\linewidth}
  \includegraphics[width=.9\linewidth]{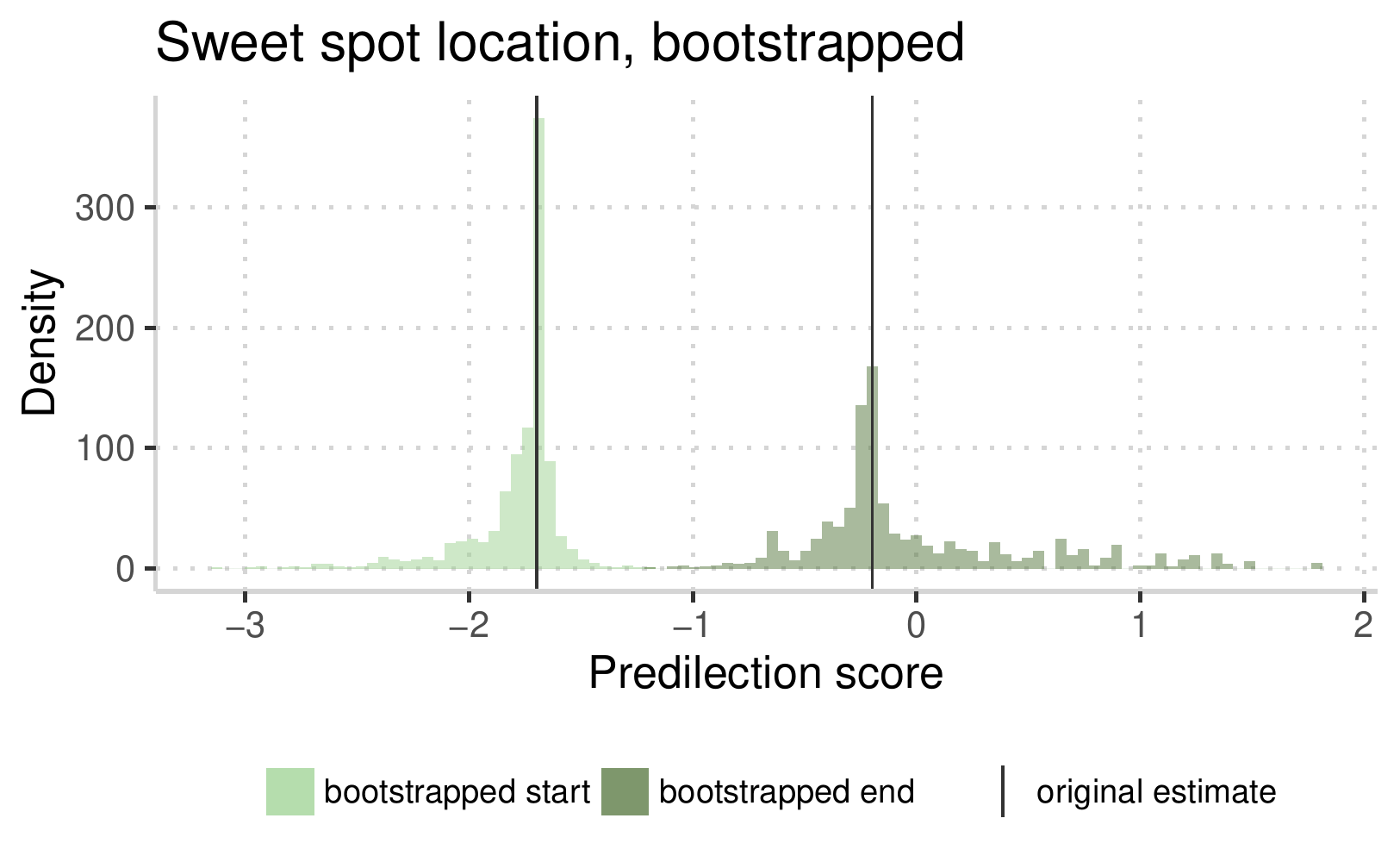}
\end{subfigure}
\caption{Visualizations from the de-biasing parametric bootstrap on the AQUAMAT data. On the left, we visualize bias: we overestimate the CATE by $0.003$. On the right, we visualize uncertainty around the location of the sweet spot.}
\label{fig:biashistogram}
\end{figure}

\begin{example*}
On the AQUAMAT data, we found $\hat{\tau} = 0.123$. Our bootstrap estimate was $\hat{\tau}_\text{boot} = 0.126$, so we overestimated by $0.003$ on average. We adjust our estimate down to $\hat{\tau}_\text{corrected} = 0.120$ (Figure~\ref{fig:biashistogram}).
\end{example*}

\subsection{More results on real data}

\begin{figure}[H]
  \centering
  \includegraphics[width=.6\linewidth]{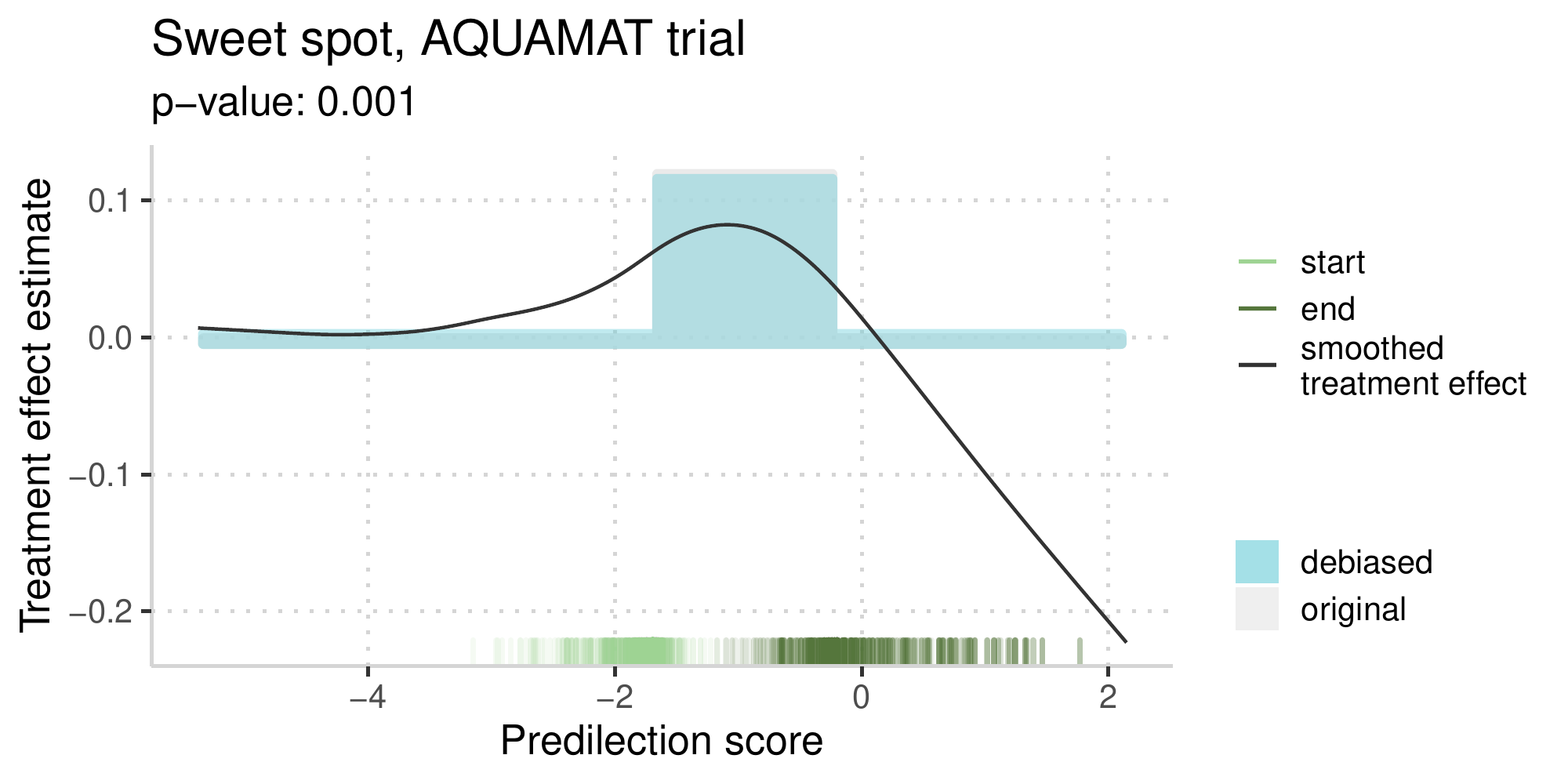}
\caption{Sweet spot found on the AQUAMAT data.}
\label{fig:final}
\end{figure}

We summarise our results on the AQUAMAT data in a single image (Figure~\ref{fig:final}). Our results include the original and debiased estimates of the CATE inside and outside the sweet spot, together with our bootstrapped distributions of the start and end index of the sweet spot. We also visualize the smoothed treatment effect as a function of predilection score.

On the AQUAMAT data, we compare our method to the Gompertz fit in \cite{redelmeier2020approach}, the causal forest in \cite{athey2019generalized}, and the reference class approach in \cite{watson2020graphing}. To compute a $p$-value for the Gompertz method, we use the bootstrap as defined in Section~\ref{section:pvalue}. 

\begin{figure}[H]
  \centering
  \includegraphics[width=.7\linewidth]{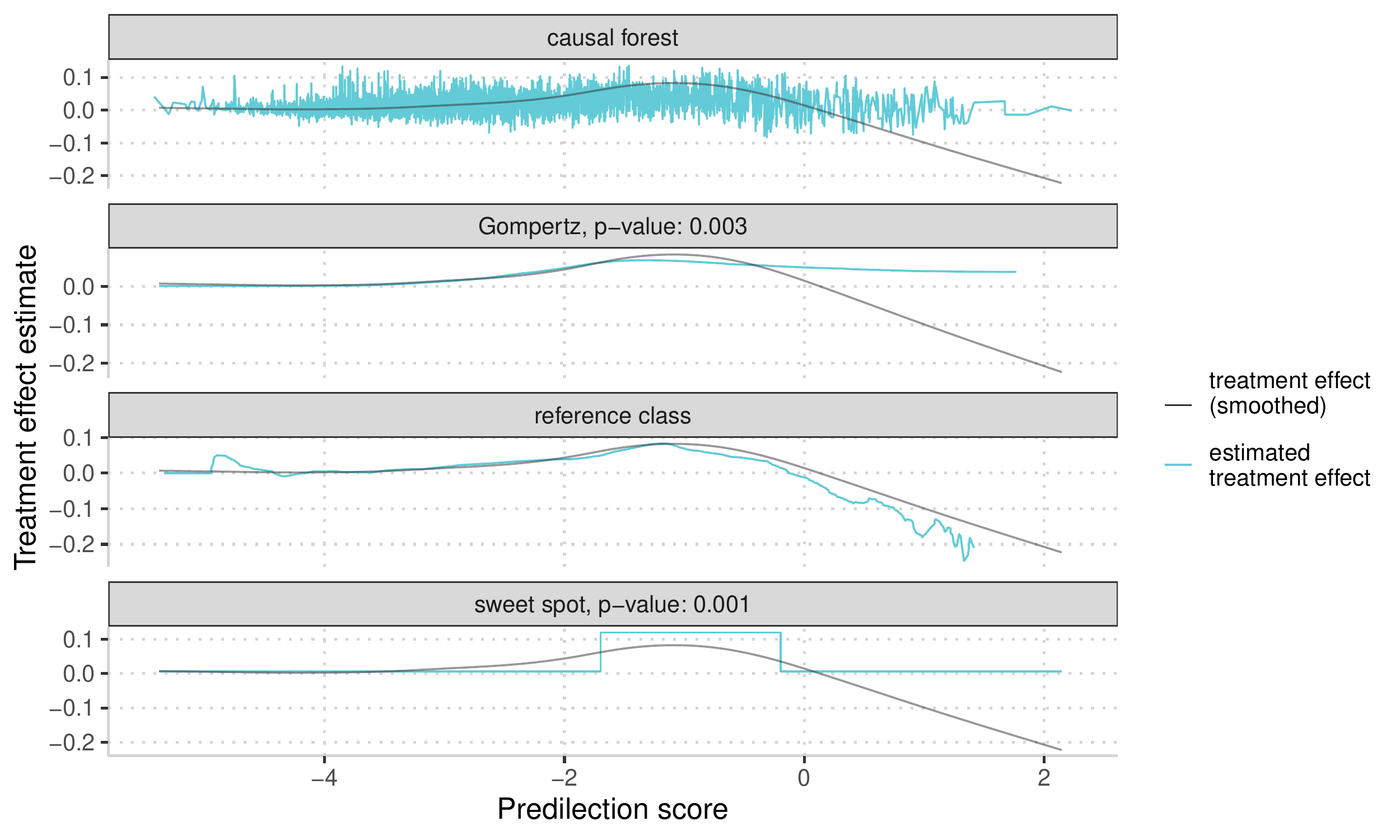}
\caption{Comparison of methods on the AQUAMAT data.}
\label{fig:comparison}
\end{figure}

We also illustrate our method on the SEAQUAMAT trial as in \cite{watson2020machine}, which compared quinine to artesunate for the treatment of severe malaria in Asian adults. The superiority of artesunate for severe malaria is now well established~\cite{white2014451}, and in this retrospective analysis, we consider artesunate to be standard of care. We follow the data preparation and experimental setup in \cite{watson2020machine}, and our finding agrees with theirs: we fail to reject the null hypothesis that there is no range of illness severity for which quinine is superior. 

\begin{figure}[H]
\centering
  \includegraphics[width=.7\linewidth]{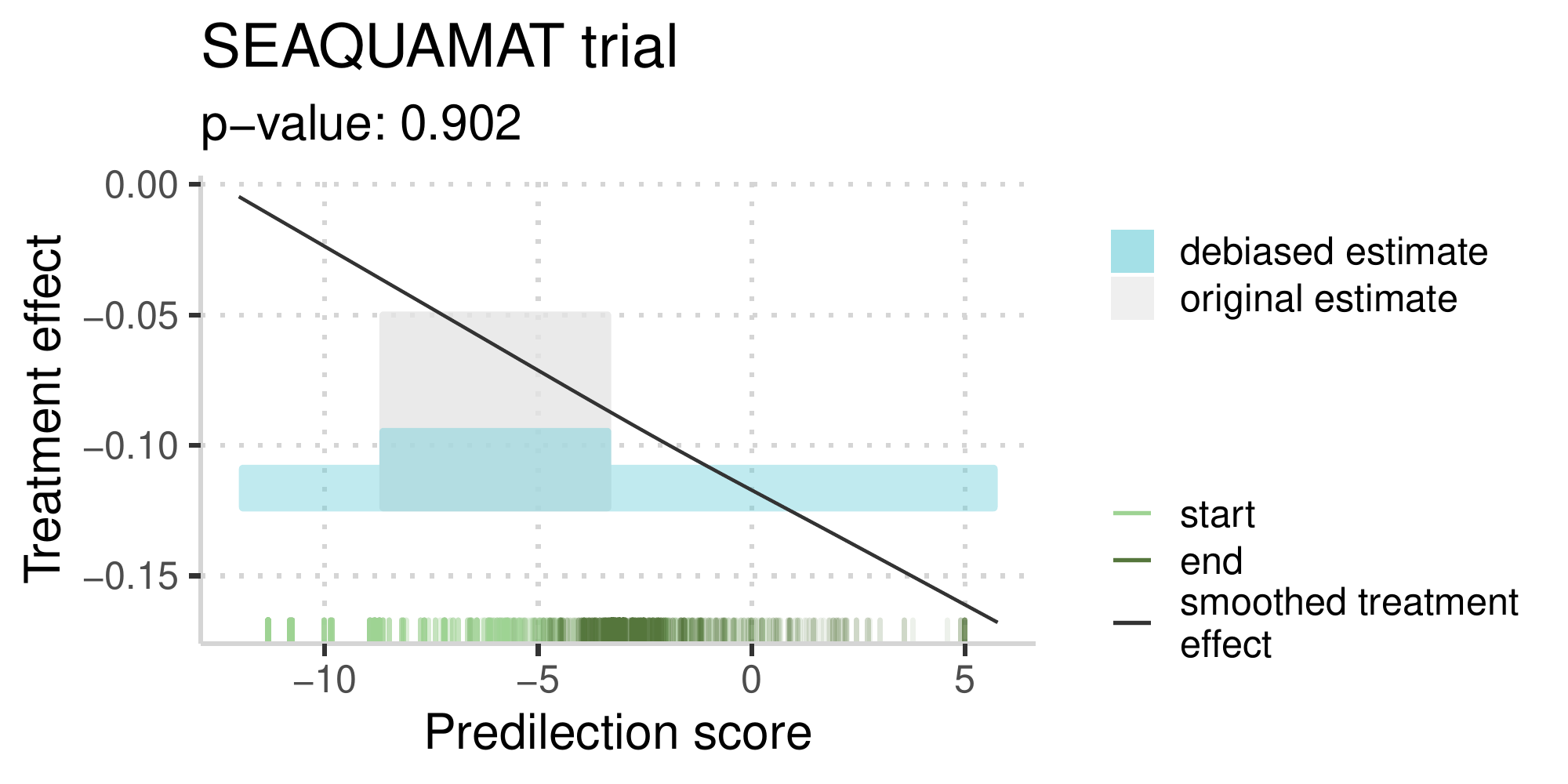}
\caption{No sweet spot found on the SEAQUAMAT trial data.}
\label{fig:seaquamat}
\end{figure}

\section{Simulation studies}
\subsection{Type I error}
\label{section:type1error}
\begin{figure}[H]
  \centering
  \includegraphics[width=.5\linewidth]{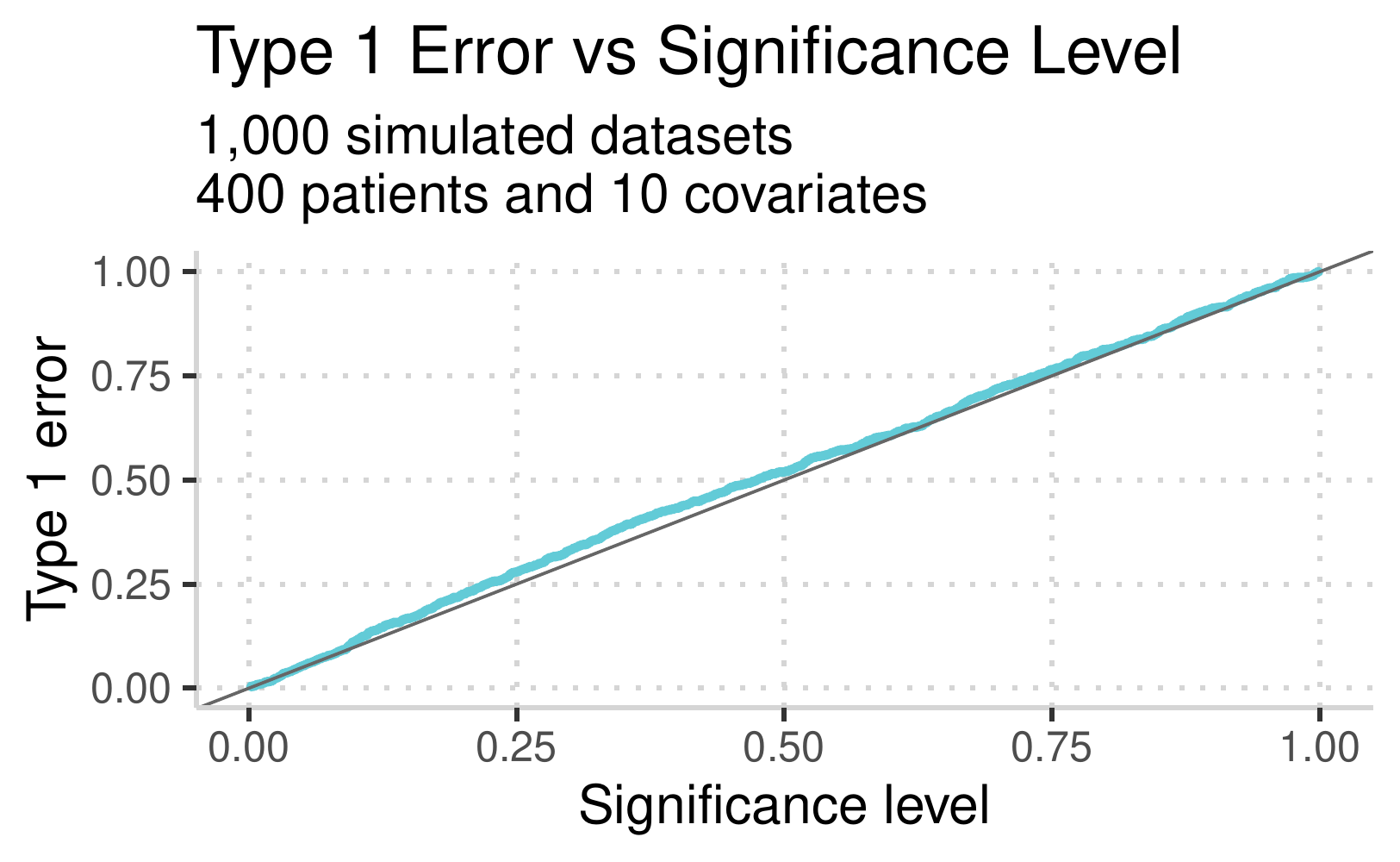}
  \caption{Type I error on simulated data.}
  \label{fig:type1}
\end{figure}

To compute the type I error of our method, we simulate randomized trial data with no sweet spot. In each of our $\num{1000}$ simulations, covariates for $400$ patients are drawn from a standard multivariate normal distribution in $10$ dimensions, and patients are assigned to receive treatment with probability $0.5$. The probability of a negative outcome is determined by a logistic model with coefficients drawn from a standard multivariate normal distribution in $10$ dimensions and a normally distributed error term. For patients who receive treatment, this probability is lowered by $0.05$ (the treatment effect). On our simulated data, we find our method to be well-calibrated, and this is illustrated in Figure~\ref{fig:type1}.

\subsection{Power}
\label{section:power}

\begin{figure}[H]
  \centering
  \includegraphics[width=.8 \linewidth]{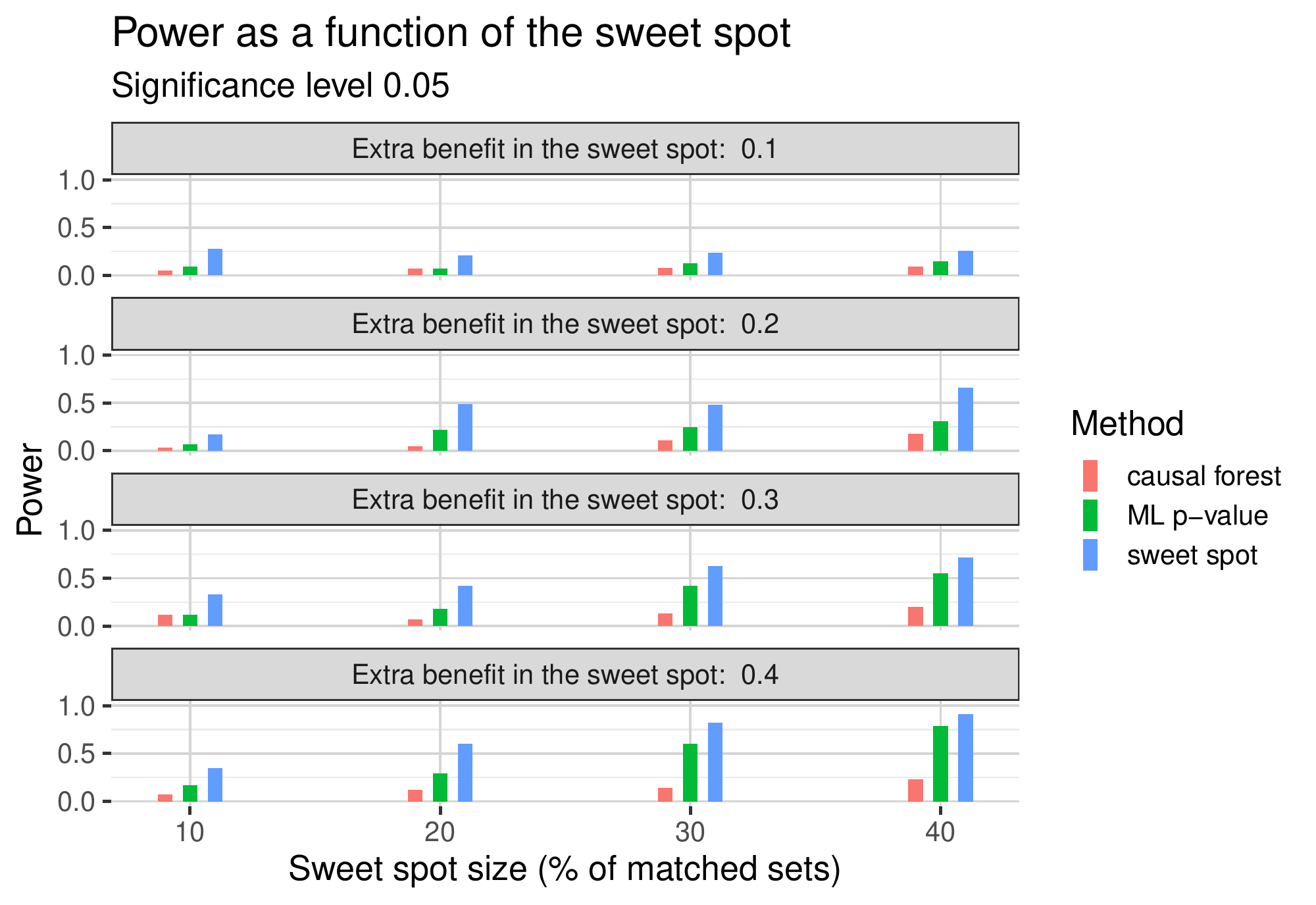}
\caption{Power as a function of sweet spot size and magnitude on simulated data. ``ML $p$-value" refers to the method in \cite{watson2020machine}.}
\label{fig:power}
\end{figure}

To compute the power of our method, we again simulate randomized trial data as in Section~\ref{section:type1error}, however now we add an extra treatment effect for patients in the middle range of illness severity. In Figure~\ref{fig:power}, we examine power along two axes: the first is the extra treatment effect in the sweet spot, and the second is the size of the sweet spot. We compare our method to a causal forest~\cite{athey2019generalized} and to the method in \cite{watson2020machine}, ``ML analysis plans for randomized controlled trials". The latter computes a $p$-value and is directly comparable to our method. To compare our method to the former, we do the following: on our simulated data, we fit a causal forest and predict outcomes, and then we perform a one-sided $t$-test that tests whether the mean predicted outcome in the sweet spot is larger than that outside the sweet spot. 

In this setting, our method has the highest power: this is expected, as we simulated data that matches our method's assumptions. This comparison is included to illustrate that all methods struggle when the sweet spot is small (covering only about $10\%$ of the data), and when there is little extra benefit in the sweet spot ($ \leq 20\%$). 

We repeat this experiment, this time defining the sweet spot location as a region defined by three of the ten covariates.  The methods in \cite{athey2019generalized} and \cite{watson2020machine} are tree based, and should be able to discover the relevant covariates. In all methods, the power is low compared to what the sweet spot method achieves in Figure~\ref{fig:power}, and the method in \cite{watson2020machine} has superior performance when the true sweet spot is large and strong. A sweet spot search over the full covariate space can only have large power when there is a large effect.

\begin{figure}[H]
  \centering
  \includegraphics[width=.8 \linewidth]{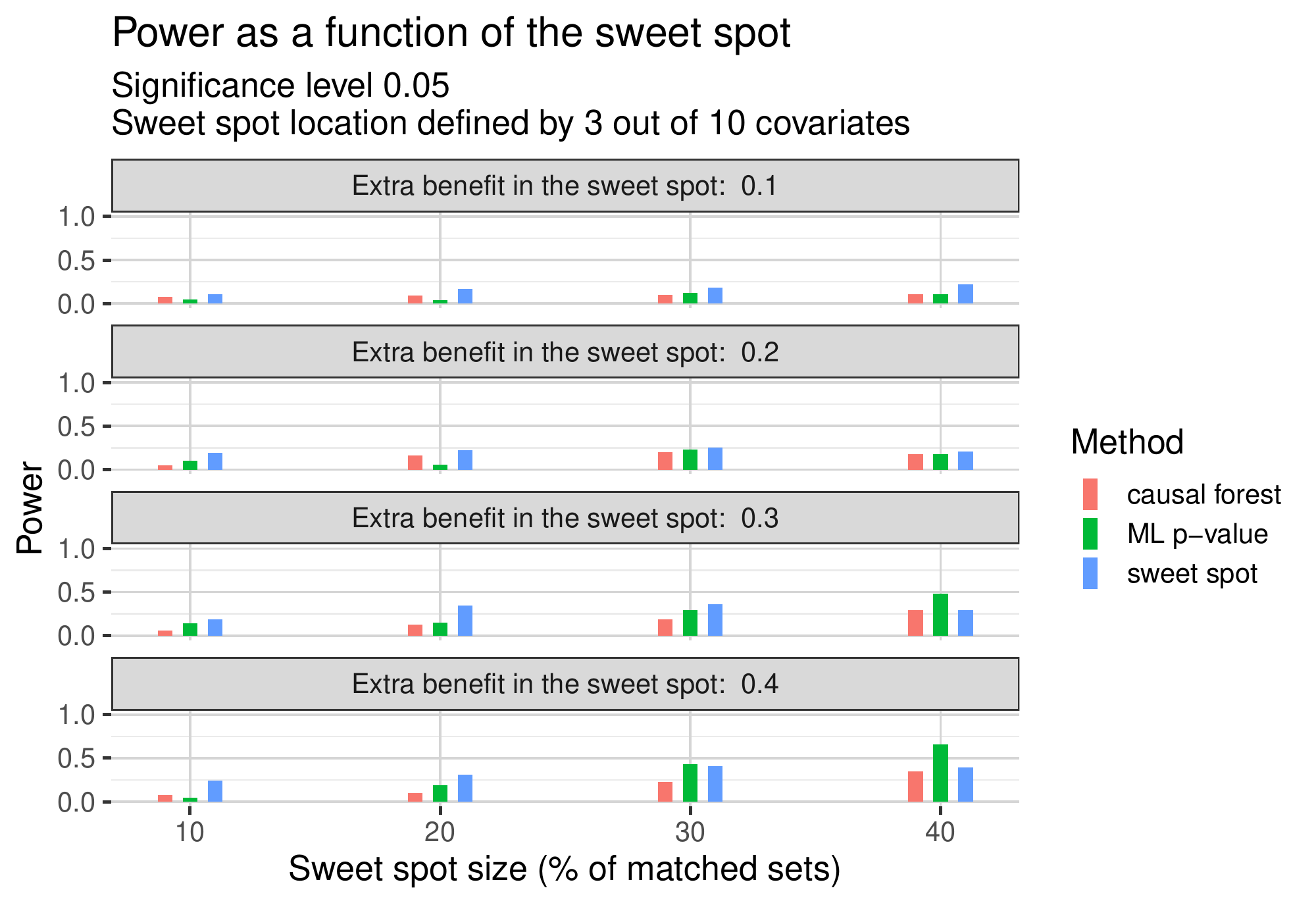}
\caption{Power as a function of sweet spot size and magnitude on simulated data, where the sweet spot is defined by three of the ten covariates. ``ML $p$-value" refers to the method in \cite{watson2020machine}.}
\label{fig:power2}
\end{figure}

\section{Further notes and discussion}

\subsection{Prevalidation}
\label{section:preval}

To compute treatment effects, we match a treated patient with a control who has similar illness severity, and we compute the difference in their outcomes. However, we do not know the \textit{true} illness severity for any patient; rather, we fit a predilection score model, and use predilection score as a proxy for illness severity. So, though we think of our matched sets as sets of patients with similar illness severity, they are more precisely sets of patients with similar \textit{predilection scores}.

Though subtle, this distinction is important. Without prevalidation, we may overfit our predilection score model to the controls. If we imagine overfitting to the extreme, a large predilection score for a control indicates only that the model knows the control had a negative outcome; a small score likewise indicates a positive outcome. Our model has no such knowledge of the future for the treated patients. 

Overfitting will have downstream effects: we use the predilection scores to pair treated and control patients. In the pairs with larger predilection scores, we overestimate the treatment effect: the control is more likely to have had a negative outcome than the treated patient. Similarly, we underestimate the treatment effect in the pairs with smaller predilection scores, as the control is more likely to have had a positive outcome. 

As a result, without prevalidating the predilection score model, we inject treatment effect heterogeneity into our data, which causes us to lose control of type I error: we are more likely to find a sweet spot when there is none. This is discussed in detail in \cite{abadie2018endogenous}.

\begin{figure}[H]
  \centering
  \includegraphics[width=.8 \linewidth]{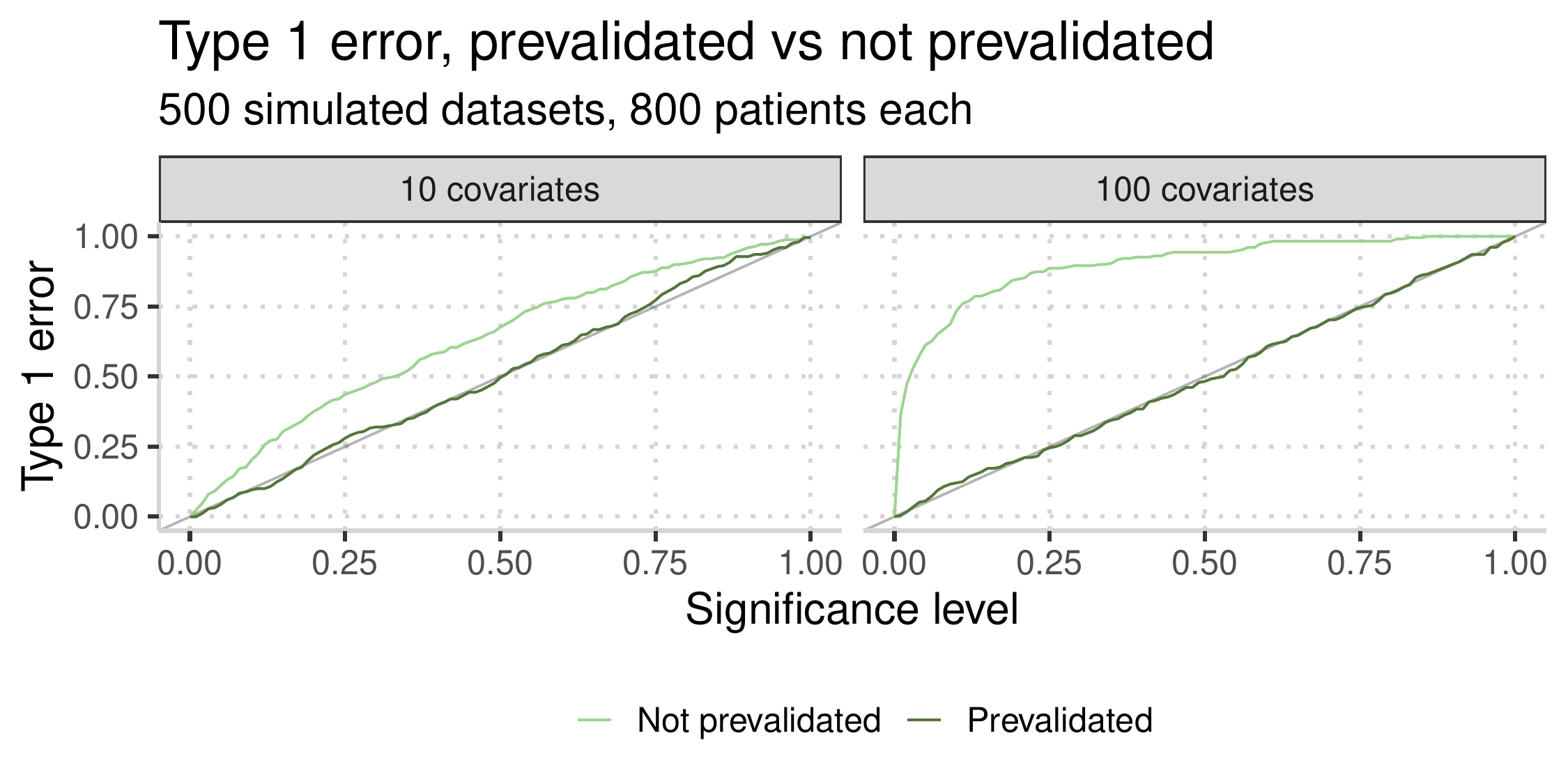}
  \caption{Type I error on simulated data, with and without prevalidation. We simulate data as in Section~\ref{section:type1error}.}  \label{fig:preval}
\end{figure}

We illustrate this in Figure~\ref{fig:preval} on simulated data, with $n=800$ trial participants and $p=10$ and $p=100$ covariates. In scenarios where we are more prone to overfitting (in this example, when $p=100$), our problem becomes more pronounced. 

\subsection{Calibration}

\begin{figure}[H]
  \centering
  \includegraphics[width=.6\linewidth]{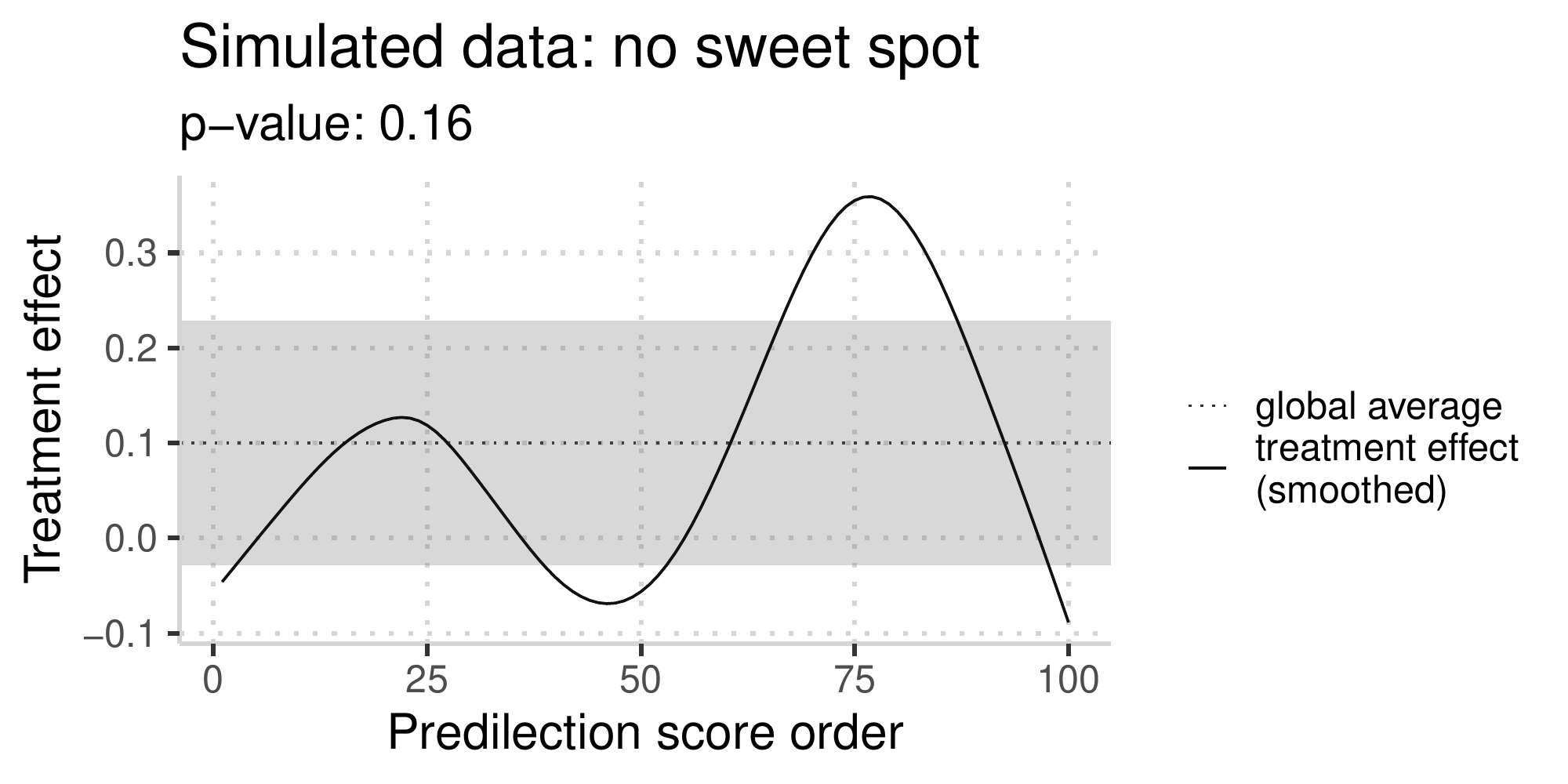}
\caption{Simulated data with no sweet spot, though it is tempting to find a sweet spot by visual inspection.}
\label{fig:calibrationtrick}
\end{figure}

Without a measure of calibration, it can be tempting to erroneously find a sweet spot. In Figure~\ref{fig:calibrationtrick}, we show a sample drawn from a data generating process with no sweet spot; by visual inspection, however, it is tempting to identify a sweet spot over the range of the $70^{\text{th}} - 84^{\text{th}}$ predilection scores. Our permutation test $p$-value is $0.160$; it correctly finds no sweet spot.

%
%

\section{Discussion}
The idea behind our method is simple: identify the range of illness severity where treatment benefit is maximized, estimate the benefit inside and outside this range, and test the null hypothesis that there is no treatment effect heterogeneity. There are existing methods for modeling treatment effect heterogeneity in randomized trials: some model the treatment effect as a function of covariates, and others identify the presence of treatment effect heterogeneity. Our method is unique as it does both, and our results are straightforward to visualize and interpret. Further, our method has a natural extension to multi-arm trials: treatments can be compared pairwise (as we have illustrated here with a treated and control group), or treatments may be compared to the group of all other treatments.

When the trial has a survival endpoint where patients may be right-censored (as in \cite{redelmeier2020approach}), estimating the treatment effect is complicated by censoring. We do not know the true outcome for all patients, and it is less obvious how to directly compare outcomes within a matched pair. This is an open area for future research.

In this paper, we exploited the relationship between treatment effect and illness severity. When the treatment effect is independent of illness severity (or it cannot be expressed by our predilection score model), our method will not find the sweet spot. In principle, another measure could be used in place of illness severity, though this is advisable only when there is a natural choice for a particular dataset. Simulations in Section~\ref{section:power} show that identifying small or weak sweet spots remains an open challenge.

\subsection{Acknowledgements}
The authors thank Lu Tian, Trevor Hastie, Alejandro Schuler, Rocky Aikens, and Stephen Pfohl for helpful discussions, and James Watson and Chris Holmes for data access. 

\begin{appendices}
\section{Finding the sweet spot}
We can speed up our computation of $Z(i, j) = \sum_{k=i}^j t_k - \frac{j-i+1}{n}  \sum_{k=1}^n t_k.$ by vectorizing. For each $k < n$, we can simultaneously compute the vector of values of $Z(i, j)$ that satisfy $j-i=k$. First, we compute the cumulative sum of $\mathbf{t}$, denoted $\mathbf{t}^*$, such that  $t_w^*= \sum_{k=1}^w t_w$. We then compute: \[ \mathbf{Z}(k) =  \{ t^*_{k+1}, t^*_{k+2}, \dots, t^*_{n} \}  - \{ t^*_1, t^*_2, \dots, t^*_{n-k} \} - k \frac{t^*_n}{n}. \] 

For very large studies, we can conserve computing time by choosing to consider only e.g. ranges where $j-i$ is even. We may also restrict to ranges within a minimum and maximum size: for example, further study of the treatment may be practical only if the sweet spot covers at least $10\%$ of the patients in the trial. 

\end{appendices}

\bibliography{main}{}

\end{document}